\documentclass[12pt]{article}

\usepackage{amssymb,amsmath,bm,bbm}
\usepackage{cite}
\usepackage{color}
\usepackage{slashed}
\usepackage{graphicx}

\allowdisplaybreaks[2]

\newcommand{\TeV}{\,\mathrm{TeV}}
\newcommand{\GeV}{\,\mathrm{GeV}}

\newcommand{\vev}[1]{\left\langle #1\right\rangle}

\newcommand{\eq}[1]{eq.~(\ref{eq:#1})}

\newcommand{\eqs}[1]{eqs.~(\ref{eq:#1})}

\newcommand{\nohyphens}%
        {\hyphenpenalty=10000\exhyphenpenalty=10000\relax}

\DeclareMathOperator{\tr}{Tr}

\newcommand{\GSM}{G_\text{SM}}

\newcommand{\ld}{\mathbf{l}}
\newcommand{\dd}{\mathbf{d^c}}

\newcommand{\Nd}{\mathbf{N^c}}
\newcommand{\preprintnumbera}{SISSA 41/2011/EP}
\newcommand{\preprintnumberb}{CP3-Origins-2011-026}
\newcommand{\preprintnumberc}{TUM-HEP-815/11}

\newlength{\myem}
\newcommand{\sep}[1]{#1}
\newcounter{mysubequation}[equation]
\renewcommand{\themysubequation}{\alph{mysubequation}}
\newcommand{\mytag}{\stepcounter{mysubequation}%
\tag{\theequation\protect\sep{\themysubequation}}}
\newcommand{\globallabel}[1]{\refstepcounter{equation}\label{#1}}

\begin{document}
\begin{titlepage}
\hspace*{\fill}
\begin{tabular}[t]{r}\preprintnumbera\end{tabular}

\hspace*{\fill}
\begin{tabular}[t]{r}\preprintnumberb\end{tabular}

\hspace*{\fill}
\begin{tabular}[t]{r}\preprintnumberc\end{tabular}

\vspace*{-1.0truecm}
\begin{flushright}
\end{flushright}

\vspace{0.4truecm}

\begin{center}
\boldmath

{\Large\textbf{Extended Tree-Level Gauge Mediation}}			

\unboldmath
\end{center}

\vspace{0.4truecm}

\begin{center}
{\bf Maurizio Monaco$^a$, Marco Nardecchia$^b$, \\
Andrea Romanino$^a$, Robert Ziegler$^{c,d}$
}
\vspace{0.4truecm}

{\footnotesize
$^a${\sl SISSA/ISAS and INFN, I–34151 Trieste, Italy}\vspace{0.2truecm}

$^b${\sl Centre for Particle Physics Phenomenology, CP3 -Origins, \\
University of Southern Denmark, Campusvej 55, \\ DK-5230 Odense M, Denmark.}\vspace{0.2truecm}

$^c${\sl Physik Department, Technische Universit\"at M\"unchen,
James-Franck-Stra{\ss}e, \\D-85748 Garching, Germany}\vspace{0.2truecm}

$^d${\sl TUM-IAS, Technische Universit\"at M\"unchen,  Lichtenbergstr.~2A,\\ D-85748 Garching, Germany \vspace{0.2truecm}}

}
\end{center}

\vspace{0.4cm}
\begin{abstract}
\noindent
Tree-level gauge mediation (TGM) is a scenario of SUSY breaking in which the tree-level exchange of heavy (possibly GUT) vector fields generates flavor-universal sfermion masses. In this work we extend this framework to the case of $E_6$ that is the natural extension of the minimal case studied so far. Despite the number of possible $E_6$ subgroups containing $\GSM$ is large (we list all rank 6 subgroups), there are only three different cases corresponding to  the number of vector messengers.
As a robust prediction we find that sfermion masses are SU(5) invariant at the GUT scale, even if the gauge group does not contain SU(5). If SUSY breaking is mediated purely by the U(1) generator that commutes with SO(10) we obtain universal sfermion masses and thus can derive the CMSSM boundary conditions in a novel scenario.

\end{abstract}
\end{titlepage}

\section{Introduction}
One of the most popular candidates for new physics at the electroweak scale is the Minimal Supersymmetric Standard Model (MSSM), tested in this very moment at the LHC. Since the bulk of its parameter space is made up by soft SUSY breaking terms, a model for SUSY breaking (and its mediation to the MSSM) is crucial in order to make definite predictions. While in popular models SUSY breaking is brought to the MSSM via gravitational~\cite{GravMSB} or SM gauge interactions at loop-level~\cite{GMSB}, we recently proposed a new framework in which SUSY breaking is communicated through new gauge interactions at tree-level~\cite{TGM1,TGM2}. We showed that this possibility is not only viable (despite the familiar arguments against tree-level SUSY breaking), but also solves the supersymmetric FCNC problem and, in its simplest SO(10) implementation, leads to peculiar relations among sfermion masses that make this scenario testable. In this paper, we want to go beyond the minimal model and analyze which of its phenomenological features persist and whether we can obtain new predictions for sfermion mass ratios. 

\begin{figure}[ht]
\begin{center}
\includegraphics[width=0.50\textwidth]{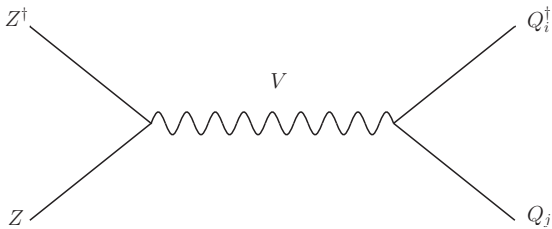}
\end{center}
\caption{Tree-level supergraph inducing sfermion masses.}
\label{Fig1}
\end{figure}

Let us shortly review the basis mechanism of tree-level gauge mediation (TGM). In this framework sfermion masses arise from an s-channel exchange of a heavy vector superfield $V$ as in Fig.~\ref{Fig1}, where $Q_i$ denote the MSSM fields and $Z$ are fields that acquire SUSY breaking F-term vevs.
Since $Z$ has to be a SM singlet, $V$ must be a SM singlet as well. The above diagram induces sfermion masses given by 
\begin{equation}
\label{mintro}
\tilde m^2_{ij} = 2 g^2 (T_a)_{ij} (M^2_V)^{-1}_{ab} F^\dagger_0 T_b F_0 ,
\end{equation}
where $a,b$ run over the SM singlet generators, $F_0$ collectively denote the F-terms vevs and $M^2_V$ is the heavy gauge boson mass matrix. Alternatively one can think of sfermion masses to arise from a D-term vev of $V$ that is induced by the F-term vevs according to
\begin{align}
\vev{D_a} = - 2 g  (M^2_V)^{-1}_{ab} F^\dagger_0 T_a F_0. 
\end{align}
Gaugino masses arise at 1-loop as in ordinary gauge mediation from coupling the above F-term vevs to heavy chiral fields with SM quantum numbers. Such heavy fields are naturally present in this scenario as they are required by the mass sum rule. The naively expected loop hierarchy between sfermion and gaugino masses typically gets reduced by various effects, e.g. the charges that enter sfermion masses. A slight hierarchy is actually desirable since it implies that the tree-level contribution to sfermion masses dominates over the two-loop contribution from ordinary gauge mediation.

Since sfermion masses arise from D-terms of new gauge fields that are SM singlets, we have to consider gauge groups with rank $\geq 5$. While such fields can certainly be present in generic extensions of the SM group, e.g. of the form $\GSM \times {\rm U(1)}'$,  a natural motivation of their presence is in the context of grand unified theories (GUTs).  In earlier works~\cite{TGM1,TGM2} 
the minimal case of rank 5 GUT extensions of the SM has been considered. Among the few, SO(10) is the obvious option, as it allows to embed the quantum numbers of a whole fermion family (plus a gauge singlet) into its simplest chiral irreducible representation, the spinorial 16. SO(10) contains only one new SM singlet gauge fields, giving rise to a particularly predictive supersymmetry breaking spectrum. In this paper we want to extend our analysis to the case of non-minimal rank 6 GUT extensions. The natural choice of the GUT group is in this case $E_6$, which, besides being independently well motivated and widely studied in the literature~\cite{E6}, is strongly motivated by TGM. In the context of SO(10) the chiral superfield spectrum needed for TGM to work contains three families of $16+10$ representations, which, together with an SO(10) singlet, form precisely the fundamental of $E_6$: $27 = 16+10+1$. The $E_6$ vector spectrum contains four new SM singlets and thus allows for a variety of possibilities for combining the corresponding D-terms. To be general, we will consider the possibility that part of $E_6$ is broken by boundary conditions in the context of extra dimensions~\cite{Heb}, so that we will deal with an effective theory below the compactification scale with a gauge group that is a rank 5 or rank 6 subgroup $G$ of $E_6$. As far as the tree-level sfermion mass prediction is concerned, what matter are the structure and the breaking of the SM singlet generators by scalar and F-term vevs, inducing D-terms for the corresponding vector superfields. Their number is either one, two or four and we will study the three cases in the next sections. Although these generators are SM singlets and do not contribute to the running of the SM gauge couplings at one loop, we will assume throughout this paper that they are broken at the GUT scale and leave an analysis of low-scale TGM to a future work. 

\section{General Framework}

Let us determine the possible TGM supersymmetry breaking messengers. The gauge group we want to consider is a subgroup $G$ of $E_6$ (including $E_6$) containing the SM group $\GSM$. The messengers are a subset of the SM singlet $E_6$ vectors. In order to identify the latter, let us decompose the $E_6$ adjoint with respect to $\GSM$ and consider the embedding of the SM group $\GSM$ in $E_6$ through the maximal subgroup $\text{SO(10)}\times\text{U(1)}_{10}$. The relevant subgroup chain is
\begin{align}
\label{chain1}
E_6 & \to \text{SO(10)} \times\text{U(1)}_{10} \to \text{SU(5)} \times\text{U(1)}_{5} \times\text{U(1)}_{10} \to 
G_\text{SM} \times\text{U(1)}_{5} \times\text{U(1)}_{10},
\end{align}
and the corresponding decomposition of the $E_6$ adjoint 78 is
\globallabel{eq:chain1}
\begin{gather}
78 \to 45_0 + 16_{-3} + \overline{16}_3 + 1_0 \mytag \\[0mm]
\begin{aligned}
45_0 & \to 24_{0,0} + 10_{-4,0} + \overline{10}_{4,0} + 1_{0,0}  &1_0 & \to 1^\prime_{0,0} \nonumber \\
16_{-3} & \to \overline{5}_{-3,-3} + 10_{1,-3} + 1_{5,-3} & \overline{16}_{3} & \to 5_{3,3} + \overline{10}_{-1,3} + 1_{-5,3} .
\end{aligned} \mytag
\end{gather}
Therefore the new four SM singlets contained in $E_6$ are the fields $1_{0,0}, 1^\prime_{0,0}, 1_{5,-3} ,1_{-5,3}$, the first two corresponding to the U(1) factors U(1)$_5$ and U(1)$_{10}$. Since all these generators commute with SU(5) the sfermion masses from TGM will be SU(5) invariant (provided the embedding of MSSM fields is in full SU(5) multiplets\footnote{We will see that this is a well-motivated assumption, even if $G$ does not contain SU(5).}). This constitutes one of the main phenomenological predictions of TGM.

As for the SM matter, we will consider an embedding in irreps of $G$ that arises from the fundamental representation $27$ of $E_6$. Under the subgroup chain in eq.~(\ref{chain1}) the 27 decomposes as
\begin{gather}
27 \to 16_1 + 10_{-2} + 1_4 \mytag \\
16_1  \to \overline{5}_{-3,1} + 10_{1,1} + 1_{5,1} \quad 10_{-2} \to \overline{5}_{2,-2} + 5_{-2,-2}  1_4  \to 1_{0,4}. \mytag
\end{gather}
We therefore have some freedom to embed the SM fields, namely we can choose whether to embed $d^c,l$ into $\overline{5}_{-3,1}$ or $\overline{5}_{2,-2}$ (or a linear superposition of both). The choice will be dictated by the requirement that the sfermion masses from TGM are positive. Moreover we only want to consider  ``pure'' embeddings of the MSSM matter fields in the 27 of $E_6$. By pure embedding we mean that each SM fermion multiplet can be embedded into a single irreducible representation of the gauge group, and the representation is the same (or equivalent) for the three families. This assumption of pure embeddings is crucial to obtain flavor universal sfermion masses (since there is no flavor problem with Higgs soft masses we allow mixed embeddings for the MSSM Higgs multiplets). 

In  order to break gauge symmetry and supersymmetry we will need scalar and F-term vevs of SM singlet fields. For this purpose we introduce a certain number of ``higgs" fields, which are distinguished from the ``matter" fields by means of a symmetry like matter or R-parity. The SM components of the matter will be denoted by small letter and the components of higgs fields by capital letters. We will consider only singlets contained in the $27,\overline{27}$ and 78 of $E_6$, which are $1_{0,4}, 1_{5,1},1_{-5,3}$ + conjugated $= N^{c \prime}, N^c, S'_+, \overline{N}^{c \prime}, \overline{N}^c, S'_-$.

As a gauge group we will consider not only $E_6$ but also a generic rank five or rank six subgroup $G$ of $E_6$ which contains $\GSM$. This is because we want to consider the possibility that part of $E_6$ is broken (to $G$) through boundary conditions in the context of extra-dimensional GUT models~\cite{Heb}. The model is in this case supposed to describe  the effective theory below the compactification scale. Without loss of generality  we can assume that $G\supseteq G_\text{SM} \times\text{U(1)}_{X} \equiv G_\text{min}$, where the U(1)$_X$ is a generic linear combination of the two U(1)s appear in the subgroup chain in eq.~(\ref{chain1}). The SM singlet generators contained in $G$ can be either just the linear combination U(1)$_X$, both U(1)$_5$ and U(1)$_{10}$ or all four singlets $1_{0,0}, 1^\prime_{0,0}, 1_{5,-3} ,1_{-5,3}$ (which form a $\text{U(1)}^\prime \times \text{SU(2)}^\prime$ subgroup of $E_6$). We now analyze the three possibilities in this order. 

\section{One Messenger Case: $G \supset$ U(1)$_X$}
\label{sec1}
We will start with the simplest case in which there is only one SM singlet generator in $G$, corresponding to a U(1)$_X$ subgroup. We assume that one can choose suitable boundary conditions such that this generator is given as general linear combination of the normalized generators $\hat{t}_{5,10}$ 
\begin{equation}
\hat{t}_X  \equiv \sin{\theta_X} \hat{t}_5 + \cos{\theta_X} \hat{t}_{10} \qquad \theta_X  \in \left[ {0, \pi} \right].
\end{equation}
Sfermion masses arise from the breaking of this generator by scalar and F-term vevs according to eq.~(\ref{mintro}).  The dependence on these vevs can be parametrized by a single real parameter $m_X^2$, whose expression in terms of the vevs can be found in Appendix A. We obtain for the sfermion mass of the sfermion $f$ with X-charge $X_f$
\begin{align}
m^2_f  = X_f m^2_X, 
\end{align}
so that the sfermion masses of the candidate matter fields in the $10_{1,1}, \overline{5}_{-3,1}, \overline{5}_{2,-2}$ are given by
\begin{align}
m^2(\overline{5}_{-3,1}) &= (-3 \hat{s}_X + \hat{c}_X) m^2_X \mytag \\
m^2(10_{1,1}) &= (\hat{s}_X + \hat{c}_X) m^2_X \mytag \\
m^2(\overline{5}_{2,-2}) &=  2(\hat{s}_X - \hat{c}_X) m^2_X, \mytag 
\end{align}
where $\hat{s}_X \equiv 1/\sqrt{40} \sin{\theta_X}$ and $\hat{c}_X \equiv 1/\sqrt{24} \cos{\theta_X}$. 
These masses satisfy the useful tree-level identity
\begin{align}
\label{sumrule}
m^2(\overline{5}_{-3,1}) + m^2(\overline{5}_{2,-2}) + m^2(10_{1,1}) = 0.
\end{align}
We now show that if we assume pure embeddings of MSSM matter and require sfermion masses to be positive, then the embeddings and therefore sfermion masses are also SU(5) invariant.  First note that the embedding of $u^c_\text{SM}$, $q_\text{SM}$, $e^c_\text{SM}$ in the 27 is unique and SU(5) invariant, namely all fields must reside in the $10_{1,1}$. As for $d^c_\text{SM}$ and $l_\text{SM}$, in principle we have two possibilities for each of them: $d^c_\text{SM} = d^c \subset \overline{5}_{-3,1}$ or $d^c_\text{SM} = d^{\prime c} \subset \overline{5}_{2,-2}$ and $l_\text{SM} = l \subset \overline{5}_{-3,1}$ or $l_\text{SM} = l' \subset \overline{5}_{2,-2}$. But the relation (\ref{sumrule}) implies that at least one of the soft terms $m^2(10_{1,1}),m^2(\overline{5}_{-3,1}),m^2(\overline{5}_{2,-2})$ must be negative. Since we require that $m^2(10_{1,1})$ is positive, either $m^2(\overline{5}_{-3,1})$ or $m^2(\overline{5}_{2,-2})$ can be positive. This means that $d^c$ and $l$ must be embedded in the same $\overline{5}$, which is $\overline{5}_{-3,1}$ if $m^2(\overline{5}_{-3,1}) > 0$ and $\overline{5}_{2,-2}$ if $m^2(\overline{5}_{2,-2}) > 0$. Therefore we have the (tree-level) prediction that sfermion soft masses are SU(5)-invariant and flavour universal: 
\begin{equation}
\label{eq:softgeneral1}
(\tilde m^2_{d^c})_{ij} = (\tilde m^2_{l})_{ij} = \tilde m^2_{\overline{5}} \delta_{ij} \qquad
(\tilde m^2_{u^c})_{ij} = (\tilde m^2_{q})_{ij} = (\tilde m^2_{e^c})_{ij} = \tilde m^2_{10} \delta_{ij} ,
\end{equation}
with generic $\tilde m^2_{\overline{5}}$ and $\tilde m^2_{10}$ depending only on $\theta_X$ and $m_X^2$.

We can consider the two simplifying cases in which either $\text{U(1)}_X = \text{U(1)}_5$ or $\text{U(1)}_X = \text{U(1)}_{10}$. In the first case we have $\hat{s}_X =1/\sqrt{40}, \hat{c}_X = 0$, which implies that we need $m_X^2>0$ and the light sfermions are $d^{c \prime}, l^\prime$ in $\overline{5}_{2,-2}$. The ratio $\tilde{m}_{10}^2/ \tilde{m}^2_{\overline{5}_{2,-2}}$ is fixed to be 1/2 and we merely reproduced the SO(10) model already considered in Ref.~\cite{TGM1}.

In the second case $\hat{c}_X =1/\sqrt{24}, \hat{s}_X = 0$ we need again $m_X^2>0$, but now the light sfermions are $d^c, l$ in $\overline{5}_{-3,1}$. We have $\tilde{m}_{10}^2 = \tilde{m}^2_{\overline{5}_{-3,1}}$ and therefore obtain SO(10) invariant sfermion masses, which follows immediately from the fact that U(1)$_{10}$ commutes with SO(10) (and the SM fermions are embedded in a single SO(10) representation).  Note that in this way we can reproduce the popular CMSSM boundary conditions for sfermion masses at the scale where U(1)$_X$ is broken (except for the Higgs masses). In this scenario they are naturally flavor-universal since they arise from (extra) gauge interactions which are universal for pure embeddings.

What regards the MSSM higgs soft masses we can have in principle a mixed embedding of $H_u$ and $H_d$ in the $27, \overline{27}$ and 78 higgs fields. That is, $H_d$ and $H_u$ can in general be a linear combination of the fields $L^{27}, L^{\prime 27}, L^{\overline{27}}, L^{78}$ and $\overline{L}^{\overline{27}}, \overline{L^\prime}^{\overline{27}}, \overline{L}^{27}, \overline{L}^{78}$, respectively. The only requirement is that the coefficient of that field that actually couples to the light MSSM matter fields is sizable, i.e. $\overline{L}^{27}$ for $H_u$ and $L^{27}$ ($L^{\prime 27}$) for $H_d$ if the light fields $d^c_{SM}, l_{SM}$ are in $\overline{5}_{2,-2}$ ($\overline{5}_{-3,1})$. The Higgs soft masses depend on the precise embedding but can range only in certain intervals that are set by the soft masses of   $L^{27}, L^{\prime 27}, L^{\overline{27}}, L^{78}$ and $\overline{L}^{\overline{27}}, \overline{L^\prime}^{\overline{27}}, \overline{L}^{27}, \overline{L}^{78}$. We find that 
\begin{align}
\label{higgssoft} 
m^2_{h_d} & \in \left[  \min\{ -3 \tilde{m}^2_{10} , - \tilde{m}^2_{\overline{5}} - \tilde{m}^2_{10} \} , \max\{ 2 \tilde{m}^2_{10}, \tilde{m}^2_{\overline{5}} \} \right] \\
m^2_{h_u} & \in \left[ \min\{ - 2 \tilde{m}^2_{10}, - \tilde{m}^2_{\overline{5}}  \} ,  \max\{ 3 \tilde{m}^2_{10} , \tilde{m}^2_{\overline{5}} + \tilde{m}^2_{10} \} \right]    ,
\end{align} 

\bigskip

In order to discuss gaugino masses we have to specify, at least in part, the superpotential. Let us start from identifying the relevant fields. We first have the chiral ``matter'' fields (defined by an appropriate assignment of a negative matter or $R$ parity) associated to subrepresentations of three $E_6$ fundamentals, $27_i$, $i=1,2,3$, and grouped of course in a set of full $G$ representations. Besides the fields of a whole SM family and two singlets, the 27 of $E_6$ contains additional 10 degrees of freedom. We have in fact two copies of the down quark and lepton fields, $d^c$, $d^{c\prime}$, $l$, $l'$ and one copy of fields with conjugate quantum numbers, $\overline{d^c}$, $\overline{l}$. This is welcome, as such extra degrees of freedom need to be (and can be easily made) heavy and, as such, they can play the role of the chiral messenger responsible of gaugino masses, as in ordinary gauge mediation\footnote{If the gauge group is not $E_6$, or it does not contain SU(2)$'$ (see below), those extra components could actually be absent. We are obviously not interested in such a case.}. Let us see how they get heavy. 

As the candidate chiral messengers have different charges under U(1)$_X$, a mass term for them can only come from the vev of a SM singlet breaking U(1)$_X$. In particular, the only possibility is to use the  $N^c$, ${N^{c \prime}}$ and $\overline{N^c}$, $\overline{N^{c \prime}}$ contained in ``Higgs'' $27$ and $\overline{27}$. Without loss of generality, we can choose a basis in the flavour space of each of such singlets in which only one of them, say $N^c_M$, $N^{c\prime}_M$, $\overline{N^c}_M$, or $\overline{N^{c\prime}}_M$, gets a vev. Mass terms for the chiral messengers then arise from the following superpotential interactions
\begin{align}
\label{eq:messengermass2}
(h^l_M)_{ij} \overline{l}_i l_j N^c_M + (h^d_M)_{ij} \overline{d^c}_i d^c_j N^c_M + 
(h^{\prime l}_M)_{ij} l'_i \overline{l}_j {N^{c\prime}_M}  + (h^{\prime d}_M)_{ij} d'^c_i \overline{d^c}_j {N^{c\prime}_M}   .
\end{align}
The couplings in the superpotential terms above can be related to each other and to other superpotential couplings by gauge invariance, depending on the choice of $G$. 

Assuming that all the couplings are non-vanishing, we need a scalar vev either for $N^c_M$ or ${N^{c\prime}_M}$, but not for both, in order to avoid mixed embeddings. In order to generate gaugino masses, the fields that get a heavy mass term must also couple to supersymmetry breaking (but not the light ones, in order to avoid negative contributions to sfermion masses). This can again be achieved only by coupling them to $N^c$, ${N^{c \prime}}$, $\overline{N^c}$, $\overline{N^{c \prime}}$ singlets getting an $F$-term vev. The relevant superpotential interactions have the same form as above,
\begin{align}
\label{eq:messengermassF}
(h^l_F)_{ij} \overline{l}_i l_j N^c_F + (h^d_F)_{ij} \overline{d^c}_i d^c_j N^c_F + 
(h^{\prime l}_F)_{ij} l'_i \overline{l}_j {N^{c\prime}_F}  + (h^{\prime d}_F)_{ij} d'^c_i \overline{d^c}_j {N^{c\prime}_F}   .
\end{align}
Gauge invariance (see~\cite{TGM2}, eq. (10)) is automatically satisfied if the field getting $F$-term vev is different from the field getting scalar vev.

In summary we can distinguish two cases depending on the embedding of the light fields $d^c_{SM},l_{SM}$
\begin{itemize}
\item[] A) $N^{c\prime}_M=0$, $N^{c\prime}_F=0$, $N^c_M = M$, $N^c_F =  F\theta^2$ \quad (light sfermions are $d^{c \prime}, l^\prime$ in $\overline{5}_{2,-2}$)
\item[] B) $N^{c}_M=0$, $N^{c}_F=0$, $N^{c \prime}_M = M$, $N^{c \prime}_F = F\theta^2$ \quad  (light sfermions are $d^c,l$ in $\overline{5}_{-3,1}$). 
\end{itemize}

This gives rise to one-loop gaugino masses $M_i$ given by
\globallabel{eq:gauginos1} 
\begin{align}
M_3 &= \frac{g^2_3}{16\pi^2} \frac{F}{M} \tr \left[h_F^d \left({h_M^d}\right)^{-1} \right] \mytag \\
M_2 &= \frac{g^2_2}{16\pi^2} \frac{F}{M} \tr \left[h_F^l \left({h_M^l}\right)^{-1} \right] \mytag \\
M_1 &= \frac{g^2_1}{16\pi^2} \frac{F}{M} \tr \left[ \frac{3}{5} h_F^l \left({h_M^l}\right)^{-1} +\frac{2}{5} h_F^d \left({h_M^d}\right)^{-1}  \right] \mytag
\end{align}
for Case A, and for Case B with the replacements $h_{F,M}^d \to h_{F,M}^{\prime d}$ and $h_{F,M}^l \to h_{F,M}^{\prime l}$.

Note that the $d^c$ and $l$ contributions to gaugino masses can be split into three contributions each, corresponding to the three messenger mass eigenstates that are related to the three eigenvalues of $(h_M^d)_{ij} M$ and $(h_M^l)_{ij} M$. Each of the three contributions should be evaluated at the corresponding mass scale.  If $G\supset \text{SU(5)}$, one gets universal gaugino masses, up to corrections from non-renormalizable operators~\cite{TGMpheno,Frigerio:2008ai}.  

\section{Two Messengers Case: $G \supset$ U(1)$_5 \times$ U(1)$_{10}$}
We now consider the case with two SM singlet generators corresponding to the U(1)$_5 \times $U(1)$_{10}$ subgroup. Since the discussion of gaugino masses and Higgs soft masses is exactly same as before we will not repeat it again and restrict to tree-level sfermion masses.

The sfermion masses of the candidate matter fields in the $10_{1,1}, \overline{5}_{-3,1}, \overline{5}_{2,-2}$ depend only on their charges under U(1)$_5 \times$ U(1)$_{10}$ and the two parameters $m_5^2$ and $m_{10}^2$ that are calculated in Appendix A. We get
\begin{align}
m^2(\overline{5}_{-3,1}) &= -3 m^2_5 + m^2_{10} \mytag \\
m^2(10_{1,1}) &= m^2_5 + m^2_{10} \mytag \\
m^2(\overline{5}_{2,-2}) &= 2 m^2_5 -2 m^2_{10} \mytag 
\end{align}
with the tree-level identity
\begin{align}
\label{sumrule2}
m^2(\overline{5}_{-3,1}) + m^2(\overline{5}_{2,-2}) + m^2(10_{1,1}) = 0.
\end{align}
As in the previous section we can use this identity to show that for pure embeddings of the matter fields and positive sfermion masses we get SU(5) invariant sfermion masses. Therefore we have the (tree-level) prediction that sfermion soft masses are SU(5)-invariant and flavour universal: 
\begin{equation}
\label{eq:softgeneral2}
(\tilde m^2_{d^c})_{ij} = (\tilde m^2_{l})_{ij} = \tilde m^2_{\overline{5}} \delta_{ij} \qquad
(\tilde m^2_{u^c})_{ij} = (\tilde m^2_{q})_{ij} = (\tilde m^2_{e^c})_{ij} = \tilde m^2_{10} \delta_{ij} ,
\end{equation}
with generic $\tilde m^2_{\overline{5}}$ and $\tilde m^2_{10}$ that depend on the scalar and $F$-term vevs according to the formulae given in Appendix A. We did not find simplifying limits with definite predictions for sfermion mass ratios other than $\tilde m^2_{\overline{5}}/\tilde m^2_{10}$ = 1/2 which was considered already in Ref.~\cite{TGM1}. The ranges for the Higgs masses are the same as in section \ref{sec1}.

\section{Four Messenger Case: $G \supset$ U(1)$^\prime \times$ SU(2)$^\prime$}

Let us now consider the case in which all the four $E_6$ candidate supersymmetry breaking messengers belong to $G$. The four messengers correspond to the $E_6$ subgroup U(1)$^\prime \times$ SU(2)$'$. The SU(2)$'$ is the one appearing in the $E_6$ maximal subgroup $E_6 \supset \text{SU(6)}\times \text{SU(2)}'$ and the U(1)$'$ is the subgroup of SU(6) that commutes with SU(5), as shown in Appendix B. We denote the corresponding generators as $t'$ and $t'_a$, $a=1,2,3$. The two additional generators, with respect to the previous section, are $t'_1$ and $t'_2$, which can be combined into two complex generators $t'_\pm = (t'_1 \pm it'_2)/\sqrt{2}$, while $t'_3$ and $t'$ are linear combinations of $t_5$ and $t_{10}$ given by $t'_3 = (t_{10}-t_5)/8$ and $t' = (3 t_5 + 5 t_{10})/4$. 

The role of the SU(2)$'$ symmetry is to make the two $\overline{5}$ and the two singlets of SU(5) in the 27 of $E_6$ equivalent, i.e.\ belonging to the same SU(2)$'$ doublet. Denoting by $(a,b)_q$ the representation which transforms as $(a,b)$ under SU(5)$\times$SU(2)$'$ and has $t' = q$, we have in fact 
\begin{equation}
\label{eq:correspondence}
\overline{5}_{-3,1} + \overline{5}_{2,-2} = (\overline{5},2)_{-1}  \qquad
1_{0,4} + 1_{5,1} =  (1,2)_{5} 
\end{equation}
while the 10 and 5 of SU(5) in the 27 are SU(2)$'$ singlets and have charge $t' = 2, -4$ respectively. This makes a qualitative difference in the way sfermion masses are generated but does not alter the conclusion in \eq{softgeneral4}. 

The masses of the supersymmetry breaking messengers and the breaking of U(1)$^\prime \times $SU(2)$^\prime$ are due to the vevs of the singlets $N^{c \prime},N^c,\overline{N^{c \prime}}, \overline{N^c}$, $S'_+$, $S'_-$, as before, which are now grouped into doublets and triplets of SU(2)$'\times$U(1)$'$. As shown in Appendix A, in the presence of an arbitrary number of such representations, the masses for the sfermions in the case of the SU(2)$'$ singlets in the 27 are given by
\globallabel{eq:singlets}
\begin{align}
m^2((10,1)_2) &=  2 m^2_1 \mytag \\
m^2((5,1)_{-4}) &= -4 m^2_1.  \mytag
\end{align}
Note that the need for non-negative tree-level soft terms for the sfermions embedded in the 10 of SU(5) requires $m^2_1 \geq0$. The $\text{SU(2)}'$ doublets in the 27 can mix, and their mass matrices are given by
\globallabel{eq:doublets}
\begin{align}
m^2((\overline{5},2)_{-1}) &= 
\begin{pmatrix}
\displaystyle \frac{m^2_3}{2}-m^2_1 & 
\displaystyle \frac{m^2_+}{\sqrt{2}} \\  
\displaystyle \frac{m^2_-}{\sqrt{2}} & 
\displaystyle - \frac{m^2_3}{2}-m^2_1
\end{pmatrix} \mytag \\[2mm]
m^2((1,2)_5) &=
\begin{pmatrix}
\displaystyle \frac{m^2_3}{2}+5 m^2_{1} & 
\displaystyle \frac{m^2_+}{\sqrt{2}} \\  
\displaystyle \frac{m^2_-}{\sqrt{2}} & 
\displaystyle - \frac{m^2_3}{2}+5m^2_{1}
\end{pmatrix}. \mytag
\end{align}
The four parameters $m^2_3$, $m^2_1$, $m^2_{\pm}$ correspond to the four messengers. The first two are real, while $m^2_+ = (m^2_-)^*$. 

The MSSM masses of the sfermions that can be embedded in a 10 of SU(5) are universal and given by $2m^2_1$ at the tree level. In order to identify the masses of the MSSM sfermions that can be embedded in a $\overline{5}$ of SU(5), we have to identify the light $d^c_i$ and $l_i$ in the multiplets $(\overline{5},2)_{-1}$. In principle, the three light leptons $l^l_i$ could be superpositions of the three $t'_3 = 1/2$ lepton doublets $l_i$ and of the three $t'_3 = -1/2$ lepton doublets $l'_i$ contained in three $(\overline{5},2)_{-1}$. On the other hand, it can be shown that the natural solution of the flavour problem requires that it must be possible to identify the three light leptons with, for example, the $t'_3 = 1/2$ lepton doublets $l_i$: $l^l_i = l_i$, up to an SU(2)$'$ rotation. This is indeed what is obtained in simple models, as shown below. The three leptons turn then out to have universal soft terms proportional to $m^2_{\overline{5}} = m^2_3/2 - m^2_1$.

If the gauge group contains SU(5), the same results hold in the $d^c$ sector. If not, the three light $d^c$ are also aligned in SU(2) space, but they could in principle be oriented in a different direction. We will see that, under plausible hypotheses, this is not the case, so that we get again the prediction that the sfermion soft masses are SU(5)-invariant and flavour universal at the tree level: 
\begin{equation}
\label{eq:softgeneral4}
(\tilde m^2_{d^c})_{ij} = (\tilde m^2_{l})_{ij} = \tilde m^2_{\overline{5}} \delta_{ij} \qquad
(\tilde m^2_{u^c})_{ij} = (\tilde m^2_{q})_{ij} = (\tilde m^2_{e^c})_{ij} = \tilde m^2_{10} \delta_{ij} ,
\end{equation}
with generic $\tilde m^2_{\overline{5}}$ and $\tilde m^2_{10}$. 

The discussion of MSSM higgs soft masses is similar as before. Now $H_d$ and $H_u$ can in general be a linear combination of the doublets in $(\overline{5},2)_{-1}$,$ (\overline{5},1)_{-6}$, $(\overline{5},1)_{4}$ and $(5,2)_{1}$, $(5,1)_{6}$, $(5,1)_{-4}$ respectively. The range for the Higgs masses (for simplicity we consider the case $m^2_+=0$) is
\begin{align*}
m^2_{h_d} & \in \left[ \min \{ -6 m_1^2, \frac{m_3^2}{2} - m_1^2 , - \frac{m_3^2}{2} - m_1^2  \} ,  \max \{ \frac{m_3^2}{2} - m_1^2 , - \frac{m_3^2}{2} - m_1^2,  4 m_1^2 \}   \right] \\
m^2_{h_u} & \in \left[ \min \{ - \frac{m_3^2}{2} + m_1^2 , \frac{m_3^2}{2} + m_1^2, - 4 m_1^2  \} ,  \max \{ 6 m_1^2, - \frac{m_3^2}{2} + m_1^2 ,  \frac{m_3^2}{2} + m_1^2 \}   \right]. 
\end{align*} 

\bigskip

The presence of the SU(2)$'$ guarantees that the MSSM $l_i$ and $d^c_i$ (and the singlets $N^c$ needed to generate masses) come together with SU(2)$'$ partners $l'_i$ and $d^{c\prime}_i$ (and $N^{c\prime}$), which need to be heavy and, as such, can play the role of the chiral supersymmetry breaking messengers responsible for one-loop gaugino masses through ordinary gauge mediation mechanism. Since they must get heavy with their conjugates, the presence of the $\overline{l}_i$, $\overline{d^c_i}$ from the $27_i$ is also guaranteed. 

Let us see how they get heavy. First, let us denote the three SU(2)$'$ doublets containing the light fields as $\ld_i = ( l_i,  l^\prime_i)^T$, $\dd_i = (d^c_i, {d^\prime}^c_i)^T$. Mass terms for the extra charged matter fields can only come from superpotential interactions
\begin{equation}
\label{eq:messmass2}
(h^l_M)_{{ij}} \overline{l}_i \ld_j \mathbf{N}^c_M+ (h^d_M)_{{ij}} \overline{d^c}_i \dd_j \mathbf{N}^c_M ,
\end{equation}
where we have assumed for simplicity that only one doublet $\mathbf{N}^c_M = (N^{c\prime},  N^c)^T$ gets a vev in the scalar component. If $G\subset \text{SU(5)}$, $h^l_M = (h^d_M)^T$, up to corrections from non-renormalizable operators~\cite{TGMpheno,Frigerio:2008ai}. We can rotate without loss of generality the vev in the $N^c$ component: $\vev{\mathbf{N}^c_M} = (0, M)^T$. Then, the $l_i$ and $d^c_i$ fields automatically end up being also massless,  and the flavour problem is naturally solved. Note also that this represents an improvement with respect to the SO(10) theory studied in~\cite{TGM1,TGM2} and with respect to the 1 and 2 messenger cases studied in the previous Sections. In those cases, in fact, the possible presence of a bare mass term $\mu_{ij} \overline{l}_i l_j$ could give rise to a non-pure embedding and to flavour non-universal soft masses. In this case, such a bare mass term is forbidden by the SU(2)$'$ symmetry. 

If more than one $\Nd$ gets a vev coupled to the light fields, the flavour problem is automatically solved if, in an appropriate SU(2)$'$ basis, all those vevs lie in the $N^c$ component only. If that is the case, we can use a basis in the $\Nd$ flavour space such that only one of them gets a vev, and we can still use \eq{messmass2}. In order to avoid negative, tree-level contributions to sfermion masses from chiral superfield exchange, we need the $N^{c\prime}$ components not to get an $F$-term either. In order to generate gaugino masses, one of the $N^c$ must however take an $F$-term vev. Gauge invariance (see~\cite{TGM2}, eq. (10)) is automatically satisfied if the field getting the $F$-term vev, $\mathbf{N}^c_F$, with $\vev{\mathbf{N}^c_F} = (0,F\theta^2)^T$, is different from the one getting the scalar vev, $\mathbf{N}^c_M$. Let 
\begin{equation}
\label{eq:messmass3}
(h^{l}_F)_{{ij}} \overline{l}_i \ld_j \mathbf{N}^c_F+ (h^{d}_F)_{{ij}} \overline{d^c}_i \dd_j \mathbf{N}^c_F 
\end{equation}
be its coupling to the chiral messengers. The gaugino masses are then given by
\globallabel{eq:gauginos4} 
\begin{align}
M_3 &= \frac{g^2_3}{16\pi^2} \frac{F}{M} \tr \left[h^{d}_F (h^d_M)^{-1} \right] \mytag \\
M_2 &= \frac{g^2_2}{16\pi^2} \frac{F}{M} \tr \left[h^{l}_F (h^l_M)^{-1} \right] \mytag \\
M_1 &= \frac{g^2_1}{16\pi^2} \frac{F}{M} \tr \left[ \frac{3}{5} h^{d}_F (h^d_M)^{-1} +\frac{2}{5} h^{d}_F (h^d_M)^{-1} \right] . \mytag
\end{align}
The $d^c$ and $l$ contributions to gaugino masses can be split into 3 contributions each, corresponding to the three messenger mass eigenstates, namely to the three eigenvalues of $M_{d^c}$ and $M_l$. Each of the three contributions should be evaluated at the corresponding mass scale. 

\section{Phenomenology}
In this section we briefly comment on some general aspects of TGM phenomenology in the setup we considered. A thorough analysis of the peculiar phenomenological implications including collider signals of TGM is in progress \cite{TGMpheno}.

In TGM models sfermion masses arise at tree level, while gaugino masses arise at one loop. As mentioned, the hierarchy between gaugino and sfermion masses that one might naively expect, potentially leading to sfermions outside the reach of the LHC and to a serious fine-tuning problem, turns out to be reduced by various effects down to a mild hierarchy. The hierarchy could actually easily be fully eliminated, but a mild hierarchy is actually welcome, as it makes the ordinary 2-loop gauge mediated pollution of tree-level sfermion masses subleading, and will be assumed in the following. 

The Higgs sector parameters are not tightly related to sfermion and gaugino masses. The $\mu$ and $B\mu$ parameters are highly model dependent\footnote{For some possible implementations see~\cite{TGM2}.}, and the Higgs soft masses depend on the Higgs embedding, which is allowed to be mixed in different representation of the gauge group, as discussed above eqs.~(\ref{higgssoft}). The coefficients $X_\text{eff}$ will be conveniently taken in their ranges, while $\mu$ and $B \mu$ will be treated as free parameters and as usual traded for $M_Z$ and $\tan{\beta}$. 

Trilinear A-terms arise typically at one loop as they are generated by the exchange of heavy chiral messengers that couple directly to MSSM fields in the superpotential. Their value is model dependent, as it is controlled by unknown superpotential parameters, but it can safely neglected in a sizeable part of the parameter space~\cite{TGM1,TGM2,TGMpheno}. 

In the following we present two representative low energy spectra that can be obtained in the present framework. As a result of the previous sections we found that the main phenomenological prediction of extended TGM is that the tree level contribution to the sfermion masses is SU(5) invariant and flavor universal, and thus parametrized by two parameters $\tilde{m}^2_{\bar{5}} ,\tilde{m}^2_{10}$ which are independent in the general case. These tree level predictions for sfermion masses hold at the messenger scale where the soft terms are generated.  In order to recover the low energy spectra we have to keep into account both the finite two loop contributions from ordinary gauge mediation and the RG effects. Since sfermion masses are in our example heavier than gaugino masses, the predictions for the sfermion mass patterns are approximately preserved at low energy. One therefore expects two separated sets of sfermions grouped according to their SU(5) representation. In the graphs below, we show two illustrative spectra, one in the case $\tilde{m}_{\bar{5}}^2 > \tilde{m}_{10}^2$ and the other in the case $\tilde{m}_{\bar{5}}^2 < \tilde{m}_{10}^2$. In the specific case where $\tilde{m}_{\bar{5}}^2 = \tilde{m}_{10}^2$ as in section \ref{sec1} the spectrum we obtain is analogue to the CMSSM case with non universal Higgs masses \cite{NUHM}. The remarkable point is that, in contrast to the CMSSM case, in which universality of sfermion masses is an {\it ad-hoc} phenomenological assumption, in our extended TGM setup it follows from the fact that SUSY breaking is mediated by a heavy U(1) gauge field which universally couples to the MSSM fields. 

\begin{figure}[ht]
\begin{center}
\begin{picture}(450,300)(0,0)
\put(25,0){\vector(0,1){280}}
\put(23,10){\line(1,0){4}}
\put(6,10){\makebox(5,5){100}}
\put(23,50){\line(1,0){4}}
\put(6,50){\makebox(5,5){500}}
\put(23,100){\line(1,0){4}}
\put(6,100){\makebox(5,5){1000}}
\put(23,150){\line(1,0){4}}
\put(6,150){\makebox(5,5){1500}}
\put(23,200){\line(1,0){4}}
\put(6,200){\makebox(5,5){2000}}
\put(23,250){\line(1,0){4}}
\put(6,250){\makebox(5,5){2500}}
\put(6,280){\makebox(5,5){GeV}}
\put(118,260){\makebox(5,5){\bf (A)}}
\put(40,11.6337){\line(1,0){35}}
\put(57,16){\makebox(5,5){$h^0$}}
\put(40,219.798){\line(1,0){35}}
\put(45,211){\makebox(5,5){$H^0$}}
\put(40,219.818){\line(1,0){35}}
\put(64,211){\makebox(5,5){$A^0$}}
\put(40,219.977){\line(1,0){35}}
\put(57,225){\makebox(5,5){$H^\pm$}}
\put(80,12.3467){\line(1,0){35}}
\put(94,2){\makebox(5,5){$\chi^0_1$}}
\put(80,24.1627){\line(1,0){35}}
\put(85,32){\makebox(5,5){$\chi^0_2$}}
\put(80,201.978){\line(1,0){35}}
\put(94,191){\makebox(5,5){$\chi^0_3$}}
\put(80,202.052){\line(1,0){35}}
\put(85,209){\makebox(5,5){$\chi^0_4$}}
\put(80,24.1757){\line(1,0){35}}
\put(105,32){\makebox(5,5){$\chi^\pm_1$}}
\put(80,202.142){\line(1,0){35}}
\put(105,209){\makebox(5,5){$\chi^\pm_2$}}
\put(80,76.2523){\line(1,0){35}}
\put(94,82){\makebox(5,5){$\widetilde{g}$}}
\put(120,193.255){\line(1,0){35}}
\put(125,185){\makebox(5,5){$\nu_\tau$}}
\put(120,193.614){\line(1,0){35}}
\put(145,185){\makebox(5,5){$\tau_L$}}
\put(120,198.305){\line(1,0){35}}
\put(147,195){\makebox(50,5){$\nu_e$ $e_L$}}
\put(120,198.515){\line(1,0){35}}
\put(120,206.454){\line(1,0){35}}
\put(160,205){\makebox(5,5){$b_R$}}
\put(120,214.346){\line(1,0){35}}
\put(125,219){\makebox(30,5){$\,d_R$}}
\put(160,138.319){\line(1,0){35}}
\put(175,129){\makebox(5,5){$\tau_R$}}
\put(160,146.083){\line(1,0){35}}
\put(175,140){\makebox(5,5){$t_R$}}
\put(160,150.379){\line(1,0){35}}
\put(180,147){\makebox(50,5){$u_R$}}
\put(160,151.31){\line(1,0){35}}
\put(145,147){\makebox(5,5){$b_L$}}
\put(160,151.91){\line(1,0){35}}
\put(202,154){\makebox(5,5){$e_R$}}
\put(160,153.65){\line(1,0){35}}
\put(139,155){\makebox(30,5){$t_L$}}
\put(160,158.602){\line(1,0){35}}
\put(165,164){\makebox(30,5){$u_L$ $d_L$}}
\put(240,0){\vector(0,1){280}}
\put(238,10){\line(1,0){4}}
\put(238,50){\line(1,0){4}}
\put(238,100){\line(1,0){4}}
\put(238,150){\line(1,0){4}}
\put(238,200){\line(1,0){4}}
\put(238,250){\line(1,0){4}}
\put(340,260){\makebox(5,5){\bf (B)}}
\put(255,11.6949){\line(1,0){35}}
\put(272,16){\makebox(5,5){$h^0$}}
\put(255,237.133){\line(1,0){35}}
\put(260,229){\makebox(5,5){$H^0$}}
\put(255,237.142){\line(1,0){35}}
\put(279,229){\makebox(5,5){$A^0$}}
\put(255,237.296){\line(1,0){35}}
\put(272,244){\makebox(5,5){$H^\pm$}}
\put(295,12.35){\line(1,0){35}}
\put(309,2){\makebox(5,5){$\chi^0_1$}}
\put(295,24.16){\line(1,0){35}}
\put(300,32){\makebox(5,5){$\chi^0_2$}}
\put(295,215.7){\line(1,0){35}}
\put(309,205){\makebox(5,5){$\chi^0_3$}}
\put(295,215.8){\line(1,0){35}}
\put(300,222){\makebox(5,5){$\chi^0_4$}}
\put(295,24.17){\line(1,0){35}}
\put(320,32){\makebox(5,5){$\chi^\pm_1$}}
\put(295,215.9){\line(1,0){35}}
\put(320,222){\makebox(5,5){$\chi^\pm_2$}}
\put(295,76.19){\line(1,0){35}}
\put(309,82){\makebox(5,5){$\widetilde{g}$}}
\put(335,108.0){\line(1,0){35}}
\put(340,98){\makebox(5,5){$\nu_\tau$}}
\put(335,108.0){\line(1,0){35}}
\put(360,98){\makebox(5,5){$\tau_L$}}
\put(335,114.8){\line(1,0){35}}
\put(327,117){\makebox(50,5){$\nu_e$ $e_L$}}
\put(335,115.1){\line(1,0){35}}
\put(335,131.3){\line(1,0){35}}
\put(325,130){\makebox(5,5){$b_R$}}
\put(335,140.0){\line(1,0){35}}
\put(340,145){\makebox(30,5){$\,d_R$}}
\put(375,169.6){\line(1,0){35}}
\put(415,163){\makebox(5,5){$\tau_R$}}
\put(375,163.2){\line(1,0){35}}
\put(390,154){\makebox(5,5){$t_R$}}
\put(375,177.7){\line(1,0){35}}
\put(352,178){\makebox(5,5){$u_R$}}
\put(375,174.7){\line(1,0){35}}
\put(415,172){\makebox(5,5){$b_L$}}
\put(375,178.4){\line(1,0){35}}
\put(415,180){\makebox(5,5){$e_R$}}
\put(375,175.5){\line(1,0){35}}
\put(366,174){\makebox(5,5){$t_L$}}
\put(375,184.1){\line(1,0){35}}
\put(375,190){\makebox(30,5){$u_L$ $d_L$}}
\end{picture}
\end{center}
\caption{Overall parameters: $m = F / M = 4.5 \TeV$, $m^2_{h_u} = - 1/5 \, m^2$, $m^2_{h_d} = 3/40 \, m^2$, $\tan \beta = 30$. Case A: $m_{\bar{5}}^2 = 1 / 5 \, m^2$, $m_{\bar{5}}^2 = 2 \, m_{10}^2$ Case B: $m_{\bar{5}}^2 = 1 / 14 \, m^2$, $m_{\bar{5}}^2 = 1/2 \, m_{10}^2$.}
\label{spectrum}
\end{figure}
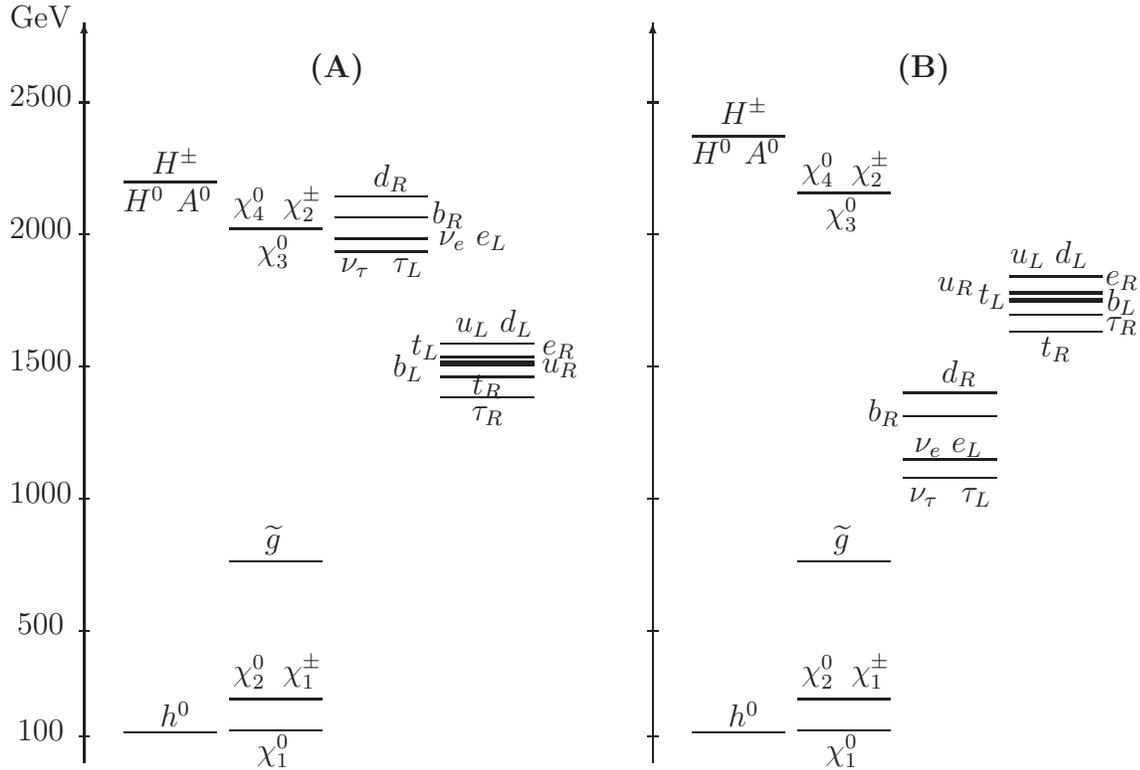

Finally, we comment about the gravitino. A general feature of TGM is the fact that the gravitino is the LSP, just as in ordinary gauge mediation. Its mass is given by 
\begin{equation}
m_{3/2} = \frac{F_0}{\sqrt{3} M_P}
\end{equation}
where $M_P = (8 \pi G_N)^{-1/2} = 2.4 \times 10^{18} \GeV$ is the reduced Planck mass and $F_0^2 = \sum_i |F_i|^2$, where the sum runs over all the fields taking $F$-term vevs. 

We note that, contrary to the minimal case, the ratios of gravitino mass and other superpartners masses are not fixed, since they depend on the specific pattern of $F$-term vevs. This happens because the $F$-terms vevs of different fields enter the gravitino mass through $F_0^2 = \sum_i |F_i|^2$, while they enter the expression for sfermion masses weighted by their charges.  A lower bound for the ratio is obtained when just one $F$-term vev is switched on. For example, in the one messenger case discussed in Sec.~\ref{sec1} one obtains
\begin{equation}
\frac{m_{3/2}^2}{m_{\widetilde{f}}^2} \gtrsim \frac{X_F}{X_f \, \sqrt{3}} \frac{M^2}{M_P^2} = 4 \times 10^{-5} \frac{X_F}{X_f} \left( \frac{M}{2 \times 10^{16} \GeV} \right)^2,
\end{equation}
where $X_f$ ($X_F$) is the charge of the sfermion (singlet breaking SUSY), and $M$ is the scalar vev responsible for U(1) breaking. On the other hand, the gravitino mass cannot be made arbitrarily large.  While gauge contributions to sfermion masses are flavour universal, gravitational ones are expected not to be. Their typical size is set by the gravitino mass, thus one has to require that (for $m_{\widetilde{f}} ^2$ around $\TeV$ scale)
\begin{equation}
\frac{m_{3/2}^2}{m_{\widetilde{f}}^2} \sim \frac{\big( m_{\widetilde{f}}^2\big)_{i \neq j}}{\big( m_{\widetilde{f}}^2\big)_{i = j}} \lesssim 10^{-4}
\end{equation}
in order to avoid flavour problems \cite{FlavorProb}.

\section{Conclusions}
In this work we have extended the framework of TGM to the case of extensions of the SM gauge group derived from $E_6$, a unified group that besides its interest for other reasons is strongly motivated by TGM. To be general, we have allowed for the possibility that part of $E_6$ is broken by boundary conditions in extra dimensions, so that we performed our analysis for an effective theory with a gauge group that is a rank 5 or rank 6 subgroup of $E_6$. Despite the large number of possible gauge groups (we gave a complete list of the rank 6 subgroups in Appendix B), we needed to study only three cases, depending on the number of vector messengers that could be one, two or four.  As a result we have found that for pure embeddings of MSSM fields we obtain SU(5) invariant (and flavor-universal) sfermion masses provided that they are positive. This feature is a pretty robust prediction of TGM that should make this scenario testable at the LHC. In the case of a rank 6 subgroup the ratio $\tilde{m}_{10}^2/\tilde{m}_5^2$ remains undetermined in the general case but can be fixed by considering special limits in the parameter space of scalar and F-term vevs to be 1/2, which is the same prediction obtained in Ref.~\cite{TGM1}. In the case of rank 5 subgroup the ratio is fixed and depends only on the specific form of the U(1) factor. If this is the U(1) subgroup of $E_6$ that commutes with SO(10) we can obtain SO(10) invariant sfermion masses. Therefore TGM offers an interesting possibility to reproduce the popular CMSSM boundary conditions for sfermion masses in a novel scenario.  In particular sfermion masses are naturally flavor-universal since, similar to ordinary gauge mediation, they arise from universal gauge interactions and in contrast to gravity mediated scenarios they can be generated at the GUT scale. 

\section*{Acknowledgements}

This work was supported in part by the Italian MIUR PRIN Project on ``Fundamental interactions in view of the Large Hadron Collider and of astro-particle physics'' (2008XM9HLM\_004), by the EU Marie Curie ITN ``UNILHC'' (PITN-GA-2009-23792), and by the ERC Advanced Grant no. 267985Ê``DaMESyFla''. R.Z. would like to thank A. Hebecker and M. Ratz for useful discussions.

\section*{Appendix A: Complete Expressions of Sfermion Masses}

In the presence of $n$ supersymmetry breaking vector messengers associated to broken generators $T^a$, $a = 1\ldots n$, sfermion masses are given by the general expression 
\begin{equation}
\label{eq:massesgeneral}
\tilde m^2_{ij} = 2 g^2 T^a_{ij} (M^2_V)^{-1}_{ab} F^\dagger_{0k} T^b_{kl} F_{0l} ,
\end{equation}
where the indices $ijkl$ denote the chiral superfields, $F_{0i}$ are the corresponding $F$-term vevs, and $M^2_V$ is the $n\times n$ messenger vector mass matrix. No matter how complicated is the Higgs mechanism giving rise to $M^2_V$ and $F_{0i}$, the sfermion masses effectively depend only on the $n$ real parameters $m^2_a \equiv 2 g^2 (M^2_V)^{-1}_{ab} F^\dagger_{0k} T^b_{kl} F_{0l}$: 
\begin{equation}
\label{eq:massesgeneral2}
\tilde m^2_{ij} = T^a_{ij} m^2_a . 
\end{equation}
The real parameters can be of course combined in complex parameters corresponding to complex generators, if needed. In each of the three cases considered in this paper, the parameters $m^2_a$ can be recovered as functions of the parameters of the model. 

In the case of one and two messengers sfermion masses arise from scalar and $F$-term vevs of the SM singlets
\begin{align}
N^{c \prime}, N^c, S'_+, \overline{N}^{c \prime}, \overline{N}^c, S'_-
\end{align}
which are understood as vectors in flavor space. We denote $(x,y)=\sum_i x_i^*y_i $, $|x|^2 = (x,x)$, where $i$ runs over the flavour indices and introduce the shorthand notation
\begin{align}
\label{eq:xnot}
x & \equiv |N^c|^2+|\overline{N^c}|^2 & y & \equiv |N^{c \prime}|^2 + |\overline{N^{c \prime}}|^2 & z & \equiv |S'_+|^2 + |S'_-|^2 \nonumber \\
f_x & = |F_{N^c}|^2- |F_{\overline{N^c}}|^2 & f_y & = |F_{N^{c \prime}}|^2 - |F_{\overline{N^{c \prime}}}|^2 & f_z & = |F_{S'_+}|^2 - |F_{S'_-}|^2,
\end{align}
where we have denoted the vevs by the same symbol used for the fields and called $F_{N^{c \prime}}, F_{N^c}, F_{\overline{N^{c \prime}}}, F_{\overline{N^c}}, F_{S'_+},F_{S'_-}$ the $F$-term vevs of $N^{c \prime},N^c,\overline{N^{c \prime}}, \overline{N^c}$, $S'_+$, $S'_-$ , respectively. 

In the one messenger case sfermion masses depend on these paramters only through a single parameter $m^2_X$ given by
\begin{align}
 m^2_X \equiv \frac{(5 \hat{s}_X + \hat{c}_X) f_x + 4 \hat{c}_X f_y + (-5 \hat{s}_X + 3 \hat{c}_X) f_z}{ (5 \hat{s}_X + \hat{c}_X)^2 x + 16 \hat{c}_X{}^2 y +  (-5 \hat{s}_X + 3 \hat{c}_X)^2 z},
\end{align}
where $\hat{s}_X \equiv 1/\sqrt{40} \sin{\theta_X}$ and $\hat{c}_X \equiv 1/\sqrt{24} \cos{\theta_X}$. 

In the two messenger case we have two parameters $m_5^2$ and $m_{10}^2$ for which we get 
\begin{align}
\begin{pmatrix}
m^2_5 \\
m^2_{10}
\end{pmatrix}  = \frac{1}{20(xy+xz+yz)}\begin{pmatrix}
f_x (4y+3z) + f_y (3z-x)-f_z(x+4y) \\
5 f_x z +  5 f_y (x+z)+ 5 f_z x
\end{pmatrix} .
\label{eq:mm5mm10}
\end{align}
Note that at least two among $x,y,z$ must be non-vanishing in order to completely break U(1)$_5 \times $U(1)$_{10}$, since a single vev would leave a linear combination of the two U(1) factors unbroken. 

In the four messenger case sfermion masses are generated by the scalar and $F$-term vevs of $n$ flavours of doublets and antidoublets and $m$ flavours of triplets
\begin{equation}
(1,2)_5 = \begin{pmatrix} N^{c \prime} \\  N^c \end{pmatrix} \quad
(1,2)_{-5} = \begin{pmatrix} \overline{N}^{c \prime} \\  \overline{N}^c \end{pmatrix} \quad (1,3)_{0} = \begin{pmatrix} S^\prime_+ \\  S^\prime_0 \\  S^\prime_- \end{pmatrix} .
\end{equation} 
In addition to \eqs{xnot} we define
\begin{gather*}
w \equiv |S^\prime_0|^2, \quad \alpha \equiv (N^{c \prime},N^c)+(\overline{N}^c , \overline{N}^{c \prime}), \quad
\beta \equiv (S^\prime_+ , S^\prime_0) + (S^\prime_0 , S^\prime_-) , \quad
\gamma \equiv (S^\prime_+,S^\prime_-),
\end{gather*}
where $\alpha,\beta,\gamma\in \mathbb{C}$, and $|\alpha| \leq \sqrt{xy}$, $|\beta| \leq \sqrt{2zw}$, $|\gamma| \leq z/2$. We use the same notation as before for the $F$-term vevs and further denote by $F_{S^\prime_0}$ the $F$-term of $S^\prime_0$. 

The sfermion masses depend on the above vevs through four parameters $m^2_+, m^2_-,m^2_3,m^2_1$ given by 
\begin{equation}
\label{eq:mmcase4}
\begin{pmatrix}
m^2_+ \\
m^2_- \\
m^2_3 \\
m^2_{1} 
\end{pmatrix} =
\begin{pmatrix}
1 & 0 & 0 & 0 \\
0 & 1 & 0 & 0 \\
0 & 0 & 1 & 0 \\
0 & 0 & 0 & 1/\sqrt{60}
\end{pmatrix} 
2 g^2 (\hat M^2_V)^{-1} 
\begin{pmatrix}
F^2_+ \\
F^2_- \\
F^2_3 \\
F^2_{1}
\end{pmatrix} ,
\end{equation}
where
\begin{equation}
\label{eq:F2}
\begin{aligned}
F^2_+ & \equiv F_0^{\dagger} \hat{T}'_+ F_0 = \frac{f_\alpha}{\sqrt{2}} - f_\beta &
F^2_- & \equiv F_0^{\dagger} \hat{T}'_-  F_0 = \left( F^2_+ \right)^* \\
F^2_3 & \equiv F_0^{\dagger} \hat{T}'_3 F_0 = f_z + \frac{f_y-f_x}{2} &
F^2_1 & \equiv F_0^{\dagger} \hat{T'} F_0 = \frac{5}{\sqrt{60}} (f_x+f_y) 
\end{aligned} 
\end{equation}
\begin{equation}
\label{eq:MMVSU2}
\hat M^2_V =g^2
\begin{pmatrix}
\displaystyle \frac{x+y+2z+4w}{2} & -2\gamma^* & -\beta^* & \displaystyle\sqrt{\frac{5}{6}} \alpha^* \\
-2\gamma & \displaystyle \frac{x+y+2z+4w}{2} & -\beta & \displaystyle\sqrt{\frac{5}{6}} \alpha \\
-\beta & -\beta ^* & \displaystyle \frac{x+y+4z}{2} & \displaystyle\frac{1}{2} \sqrt{\frac{5}{3}} (y-x) \\
\displaystyle\sqrt{\frac{5}{6}} \alpha & \displaystyle\sqrt{\frac{5}{6}} \alpha^* & \displaystyle\frac{1}{2} \sqrt{\frac{5}{3}} (y-x) & \displaystyle\frac{5}{6} (x+y)
\end{pmatrix} .
\end{equation}

\section*{Appendix B: Rank 6 Subgroups of $E_6$ containing $\GSM$}

We now provide a complete list of the rank 6 subalgebras $\mathfrak{g}$ of the $E_6$ Lie algeba containing the SM algebra. We distinguish the two ($t'_\pm\notin \mathfrak{g}$) and four ($t'_\pm\in \mathfrak{g}$) messenger cases. We will write the subalgebras as direct sums of the decomposition of the $E_6$ Lie algebra with respect to $G_\text{min} = \GSM\times\text{U(1)}_{10}\times\text{U(1)}_5$ and $G'_\text{min} = \GSM\times \text{U(1)}'\times\text{SU(2)}'$ respectively. 

The 78 decomposes as in \eq{chain1}. The $G_\text{min}$ irreducible subalgebras (besides the ones in $\mathfrak{g}_\text{min}$) can be labelled as follows: 
\begin{equation}
\label{eq:subs1}
\begin{gathered}
24_{0,0} + 1_{0,0} + 1'_{0,0} = \mathfrak{g}_\text{min} + V_{0,0} + \overline{V}_{0,0} \\
10_{-4,0} = q_{-3,1/2}+u^c_{-3,1/2}+e^c_{-3,1/2} \qquad
\overline{10}_{4,0} = \overline{q}_{3,-1/2}+\overline{u^c}_{3,-1/2}+\overline{e^c}_{3,-1/2} \\
10_{1,-3} = q_{-3,-1/2}+u^c_{-3,-1/2}+e^c_{-3,-1/2} \qquad
\overline{10}_{-1,3} = \overline{q}_{3,1/2}+\overline{u^c}_{3,1/2}+\overline{e^c}_{3,1/2} \\ 
\overline{5}_{-3,-3} = l_{-6,0} + d^c_{-6,0} \qquad
5_{3,3} = \overline{l}_{6,0} + \overline{d^c}_{6,0}  \\
1_{5,-3} = s'_{0,-1} \qquad
1_{-5,3} = s'_{0,1} .
\end{gathered}
\end{equation}
where $r_{a,b}$ denotes the subalgebra with the quantum numbers of the SM representation $r$ and with $t'=a$, $t'_3=b$ (we use $t'$ and $t'_3$ instead of $t_5$, $t_{10}$ here because it makes easier to compute commutators). $V$ denotes the (3,2,-5/6) SM representation that describes the heavy SU(5)/$G_{\rm SM}$ vectors. 

The rank 6 subalgebras $\mathfrak{g}$ of the $E_6$ Lie algeba containing the SM algebra, but not $t'_\pm$, are then
\globallabel{eq:closedsubs1}
\begin{gather}
\mathfrak{su}(5) + \mathfrak{u}(1)_5 + \mathfrak{u}(1)_{10} = \mathfrak{g}_\text{min} + V_{0,0} + \overline{V}_{0,0} \mytag \\
\mathfrak{su}(5)_f + \mathfrak{u}(1)_{5f} + \mathfrak{u}(1)_{10} = \mathfrak{g}_\text{min} + q_{-3,1/2} + \overline{q}_{3,-1/2} \mytag \\
\mathfrak{su}(4)_c + \mathfrak{su}(2)_L + \mathfrak{u}(1)_{3R} + \mathfrak{u}(1)_{10} = \mathfrak{g}_\text{min} + u^c_{-3,1/2} + \overline{u^c}_{3,-1/2} \mytag \\
\mathfrak{su}(3)_c + \mathfrak{su}(2)_L + \mathfrak{su}(2)_R + \mathfrak{u}(1)_{B-L} + \mathfrak{u}(1)_{10} = \mathfrak{g}_\text{min} + e^c_{-3,1/2} + \overline{e^c}_{3,-1/2} \mytag \\
\mathfrak{su}(3)_c + \mathfrak{su}(3)_L + \mathfrak{u}(1)'_{8} + \mathfrak{u}(1)'_{3} = \mathfrak{g}_\text{min} + l_{-6,0} + \overline{l}_{6,0} \mytag \\
\mathfrak{su}(4)_{cf} + \mathfrak{su}(2)_L + \mathfrak{u}(1)'_{3} + \mathfrak{u}(1)_{10f} = \mathfrak{g}_\text{min} + d^c_{-6,0} + \overline{d^c}_{6,0}  \mytag \\
\mathfrak{so}(10) + \mathfrak{u}(1)_{10} = \mathfrak{g}_\text{min} + (V_{0,0} + q_{-3,1/2} + u^c_{-3,1/2} + e^c_{-3,1/2} +\text{conj})  \mytag \\
\mathfrak{su}(6) + \mathfrak{u}(1)'_{3} = \mathfrak{g}_\text{min} + (V_{0,0} + l_{-6,0} + d^c_{-6,0}  +\text{conj})  \mytag \\
\mathfrak{su}(6)_f + \mathfrak{u}(1)'_{3R} = \mathfrak{g}_\text{min} + (q_{-3,1/2} + u^c_{-3,-1/2}+l_{-6,0} +\text{conj})  \mytag \\
\mathfrak{su}(5)_f + \mathfrak{su}(2)'_{R} + \mathfrak{u}(1)'_{f} = \mathfrak{g}_\text{min} + (q_{-3,1/2} + e^c_{-3,-1/2} +\text{conj})  \mytag \\
\mathfrak{su}(6)_f + \mathfrak{su}(2)'_{R} = \mathfrak{g}_\text{min} + (q_{-3,1/2} + u^c_{-3,-1/2}+l_{-6,0} +e^c_{-3,-1/2} +\text{conj})  \mytag \\
\mathfrak{su}(4)_c + \mathfrak{su}(2)_{L} + \mathfrak{su}(2)_{R} + \mathfrak{u}(1)_{10}  = \mathfrak{g}_\text{min} + (u^c_{-3,1/2} + e^c_{-3,1/2} +\text{conj})  \mytag \\
\mathfrak{su}(5)'_{fR} + \mathfrak{su}(2)_{L} + \mathfrak{u}(1)'_{fR} = \mathfrak{g}_\text{min} + (u^c_{-3,1/2} + e^c_{-3,-1/2} + d^c_{-6,0}+\text{conj})  \mytag \\
\mathfrak{su}(3)_c + \mathfrak{su}(3)_{L} + \mathfrak{su}(2)_{R} + \mathfrak{u}(1)_{R} = \mathfrak{g}_\text{min} + (e^c_{-3,1/2} + l_{-6,0} +\text{conj}),  \mytag \ \
\end{gather}
besides of course $\mathfrak{su}(3)_c + \mathfrak{su}(2)_{L} + \mathfrak{u}(1)_{Y} + \mathfrak{u}(1)_{5} + \mathfrak{u}(1)_{10} = \mathfrak{g}_\text{min}$. For the definition of the U(1) factors see Table~\ref{tableU1s}.

\begin{table}[ht]
\begin{center}
\begin{tabular}{{|c|c|c|}}
\hline
& \, {\rm Generator} \, & {\rm Definition} \\
\hline
\hline
U(1)$_{5f}$ & $t_{5f}$ & $(t_5 + 24 y)/5$ \\
U(1)$_{3R}$ & $t_{3R}$ & $(t_5 - 6 y)/10$ \\
U(1)$_{B-L}$ & $t_{B-L}$ & $(t_5 + 4 y)/5$ \\
U(1)$_8^\prime$ & $y^\prime$ & $(- 3 t_5 + 48 y - 5 t_{10})/60$ \\
U(1)$_3^\prime$ & $t_3^\prime$ & $(t_{10} - t_5)/8$ \\
U(1)$_{10f}$ & $t_{10f}$ & $(3 t_5 + 5 t_{10} + 72 y)/20$ \\
U(1)$_{3R}^\prime$ & $t^\prime_{3R}$ & $(t_5 - 5 t_{10} + 24 y)/40$ \\
U(1)$_f^\prime$ & $t^\prime_f$ & $(3 t_5 + 25 t_{10} + 72 y)/20$ \\
U(1)$_{fR}^\prime$ & $t_{fR}^\prime$ & $(- 3 t_5 + 5 t_{10} + 18 y)/5$ \\
U(1)$_R$ & $t_R$ & $(- 3 t_5 - 12 y + 5 t_{10})/30$  \\
\hline
U(1)$^\prime$ & $t^\prime$ & $(5 t_{10} + 3 t_5)/4$ \\
U(1)$_{10f}^\prime$ & $t_{10f}^\prime$ & $(3 t_5 + 5 t_{10} + 72 y)/20$ \\
U(1)$_c^\prime$ & $t_c^\prime$ & $(- 3 t_5 - 5 t_{10} + 18 y )/5$ \\
U(1)$_{8L}$ & $y_{L}$ & $( - 3 t_5 - 5 t_{10} - 12 y )/30$ \\
\hline
\end{tabular}
\caption{Definition of U(1) factors}
\label{tableU1s}
\end{center}
\end{table}

Some comments are in order. All the subgroup factors in \eqs{closedsubs1} are orthogonal. Adding a subalgebra with opposite values of $t'_3$ leads to an equivalent embedding that can be obtained from the original one by means of a SU(2)$'$ rotation flipping the sign of $t'_3$. The subalgebra $\mathfrak{su}(5)_f$ gives the flipped embedding of SU(5) in SO(10)$\subset E_6$ with the flipped U(1) generator $t_{5f}$. The ``flipped SU(4)$_c$'' subalgebra $\mathfrak{su}(4)_{cf}$ can be seen as the SU(4) subgroup of SU(6) generated by $\mathfrak{su}(3)_c + d^c_{-6,0} + \overline{d^c}_{6,0}$ and the ``flipped B-L" generator $t^f_{B-L} \equiv (t^\prime - 2 y)/5$. The flipped $\mathfrak{su}(6)_{f}$ subalgebra is spanned by $\mathfrak{su}(5)_{f} + \overline{5}_{-3,-3} + 5_{3,3} + \mathfrak{u}(1)'_{f}$. The SU(5)$'_{fR}$ subgroup is the one obtained from the unification of SU(3)$_c$ and SU(2)$'_R$ instead of SU(2)$_L$

In the case in which the gauge group contains SU(2)$'$ (i.e.\ $t'_\pm\in\mathfrak{g}$), it is convenient to decompose the $E_6$ adjoint with respect to $G'_\text{min}$. One has 
\begin{equation}
\begin{gathered}
78 \to (24,1)_0 + (5,1)_6 + (\overline{5} ,1 )_{-6} + (1,1)_0 + (1,3) + (10,2)_{-3} + (\overline{10},2)_3 \\
(24,1)_0 + (1,1)_0 + (1,3)_0 = \mathfrak{g}'_\text{min}  + (V,1)_0 + (\overline{V},1)_0 \\
(10,2)_{-3} = (q,2)_{-3} + (u^c,2)_{-3} + (e^c,2)_{-3} \qquad
(\overline{10},2)_{3} = (\overline {q},2)_{3} + (\overline {u^c},2)_{3} + (\overline {e^c},2)_{3} \\
(\overline{5},1)_{-6} = (l,1)_{-6} + (d^c,1)_{-6} \qquad
(5,1)_6 = (\overline{l},1)_6 + (\overline{d^c},1)_6 ,
\end{gathered}
\end{equation}
where $(a,b)_q$ denotes a subalgebra with quantum numbers $a$ under SU(5) (first line) of $\GSM$ (other lines), $b$ under SU(2)$'$, and $t' = q$. 

The rank 6 subalgebras $\mathfrak{g}$ of the $E_6$ Lie algeba containing the SM algebra and $t'_\pm$, are then
\globallabel{eq:closedsubs2}
\begin{gather}
\mathfrak{su}(5) + \mathfrak{u}(1)' + \mathfrak{su}(2)' = \mathfrak{g}'_\text{min} + [(V,1)_{0} + \text{conj}] \mytag \\
\mathfrak{su}(6) + \mathfrak{su}(2)' = \mathfrak{g}'_\text{min} + [(V,1)_{0} + (l,1)_{-6} + (d^c,1)_{-6} + \text{conj}] \mytag \\
\mathfrak{so}(10)'_f + \mathfrak{u}(1)'_{10f} = \mathfrak{g}'_\text{min} + [(q,2)_{-3} + (d^c,1)_{-6} + \text{conj}] \mytag \\
\mathfrak{su}(5)_c + \mathfrak{su}(2)_L + \mathfrak{u}(1)'_c = \mathfrak{g}'_\text{min} + [(u^c,2)_{-3} + \text{conj}] \mytag \\
\mathfrak{su}(6)_c + \mathfrak{su}(2)_L  = \mathfrak{g}'_\text{min} + [(u^c,2)_{-3} + (e^c,2)_{-3} + (d^c,1)_{-6} + \text{conj}] \mytag \\
\mathfrak{su}(3)_c + \mathfrak{su}(2)_L +\mathfrak{su}(3)' + \mathfrak{u}(1)_{8L}  = \mathfrak{g}'_\text{min} + [(e^c,2)_{-3} + \text{conj}] \mytag \\
\mathfrak{su}(3)_c + \mathfrak{su}(3)_L +\mathfrak{su}(3)' = \mathfrak{g}'_\text{min} + [(e^c,2)_{-3} + (l,1)_{-6} + \text{conj}] \mytag \\
\mathfrak{su}(3)_c + \mathfrak{su}(3)_L + \mathfrak{su}(2)' + \mathfrak{u}(1)'_{8} = \mathfrak{g}'_\text{min} + [l_{-6,0}  +\text{conj}] \mytag \\
\mathfrak{su}(4)_{cf} + \mathfrak{su}(2)_L + \mathfrak{su}(2)' + \mathfrak{u}(1)_{10f} = \mathfrak{g}'_\text{min} + [ d^c_{-6,0} + \text{conj}],  \mytag 
\end{gather}
besides of course $\mathfrak{e}_6$ itself. 

\newpage

\begin{table}
\caption{Decomposition of {\bf 27}}
$
\begin{array}{{|c|c|c|c||c|}}
\hline
\hline
 SU(6) \times SU(2)^\prime & SU(5) \times SU(2)^\prime \times  U(1)^\prime & SU(5) \times  U(1)_5 \times  U(1)_{10} & SM  & SO(10) \times  U(1)_{10} \\ 
\hline
({\bf 15,1}) & ({\bf 10,1})_2 & {\bf 10}_{1,1} &  q, u^c,e^c& {\bf 16}_1 \\
 & ({\bf 5,1})_{-4} &  {\bf 5}_{-2,-2} & \overline{d^c},\overline{l} & {\bf 10}_{-2} \\
& & & & \\
({\bf \overline{6},2}) & {\bf \overline{5}}_{-3,1}  & ({\bf \overline{5},2})_{-1} & d^c,l &  {\bf 16}_1 \\
& &  {\bf \overline{5}}_{2,-2} &  d^{\prime c},l^\prime & {\bf 10}_{-2} \\
& ({\bf 1,2})_5 &  {\bf 1}_{0,4} & \nu^{c \prime} &  {\bf 1}_4 \\
 & & {\bf 1}_{5,1} &  \nu^c &  {\bf 16}_1 \\
\hline
\hline
\end{array}
$
\end{table}

\begin{table}
\caption{Decomposition of {\bf 78}}
$
\begin{array}{{|c|c|c|c||c|}}
\hline
\hline
 SU(6) \times SU(2)^\prime & SU(5) \times SU(2)^\prime \times U(1)^\prime &  SU(5) \times U(1)_5 \times U(1)_{10}  & SM & SO(10) \times U(1)_{10} \\ 
\hline
({\bf 35,1}) & ({\bf 24,1})_0 &  {\bf 24}_{0,0} & & {\bf 45}_0 \\
 & ({\bf 5,1})_{6} & {\bf 5}_{3,3} &  &   {\bf \overline{16}}_3 \\
& ({\bf \overline{5},1})_{-6} & {\bf \overline{5}}_{-3,-3} & &   {\bf 16}_{-3} \\
& ({\bf 1,1})_0 & {\bf 1}_{0,0} &  s^\prime & ({\bf 1}_0,{\bf 45}_0) \\
& & & & \\
({\bf 20,2}) & ({\bf 10,2})_{-3}  & {\bf 10}_{-4,0} & & {\bf 45}_0 \\
& &  {\bf 10}_{1,-3} & &  {\bf 16}_{-3} \\
& ({\bf \overline{10},2})_{3} & {\bf \overline{10}}_{-1,3} & & {\bf \overline{16}}_3 \\
&  & {\bf \overline{10}}_{4,0} & &  {\bf 45}_0 \\
& & & & \\
({\bf 1,3}) & ({\bf 1,3})_0  & {\bf 1}_{-5,3} & s^\prime_+ &  {\bf \overline{16}}_3 \\
& &  {\bf 1}_{0,0} & s^\prime_0 &  ({\bf 1}_0,{\bf 45}_0) \\
&& {\bf 1}_{5,-3}  & s^\prime_-  &  {\bf 16}_{-3} \\
\hline
\hline
\end{array}
$
\end{table}

\end{document}